\documentstyle{mn}
\newif\ifAMStwofonts
\AMStwofontstrue

\newcommand{\plotone}[1]{\epsfxsize=1.0\hsize\epsfbox{#1}}
\newcommand{\plotnine}[1]{\center\epsfxsize=0.9\hsize\epsfbox{#1}}

\include{epsf}
\ifoldfss
  \newcommand{\rmn}[1] {{\rm #1}}

  \ifCUPmtlplainloaded \else
    \NewTextAlphabet{textbfit} {cmbxti10} {}
    \NewTextAlphabet{textbfss} {cmssbx10} {}
    \NewMathAlphabet{mathbfit} {cmbxti10} {} 
    \NewMathAlphabet{mathbfss} {cmssbx10} {} 
  \fi
  \ifAMStwofonts
    \ifCUPmtlplainloaded \else
      \NewSymbolFont{upmath} {eurm10}
      \NewSymbolFont{AMSa} {msam10}
      \NewMathSymbol{\upi}     {0}{upmath}{19}
      \NewMathSymbol{\umu}     {0}{upmath}{16}
      \NewMathSymbol{\upartial}{0}{upmath}{40}
      \NewMathSymbol{\leqslant}{3}{AMSa}{36}
      \NewMathSymbol{\geqslant}{3}{AMSa}{3E}

    \fi
  \fi
\fi 

\ifnfssone
  \newmathalphabet{\mathit}
  \addtoversion{normal}{\mathit}{cmr}{m}{it}
  \addtoversion{bold}{\mathit}{cmr}{bx}{it}
  \newcommand{\rmn}[1] {\mathrm{#1}}

  \newmathalphabet{\mathbfit} 
  \addtoversion{normal}{\mathbfit}{cmr}{bx}{it}
  \addtoversion{bold}{\mathbfit}{cmr}{bx}{it}
  \newmathalphabet{\mathbfss} 
  \addtoversion{normal}{\mathbfss}{cmss}{bx}{n}
  \addtoversion{bold}{\mathbfss}{cmss}{bx}{n}
  \ifAMStwofonts
    \ifCUPmtlplainloaded \else
      \UseAMStwoboldmath
      \makeatletter
      \new@mathgroup\upmath@group
      \define@mathgroup\mv@normal\upmath@group{eur}{m}{n}
      \define@mathgroup\mv@bold\upmath@group{eur}{b}{n}
      \edef\UPM{\hexnumber\upmath@group}
      \new@mathgroup\amsa@group
      \define@mathgroup\mv@normal\amsa@group{msa}{m}{n}
      \define@mathgroup\mv@bold\amsa@group{msa}{m}{n}
      \edef\AMSa{\hexnumber\amsa@group}
      \makeatother
      \mathchardef\upi="0\UPM19
      \mathchardef\umu="0\UPM16
      \mathchardef\upartial="0\UPM40
      \mathchardef\leqslant="3\AMSa36
      \mathchardef\geqslant="3\AMSa3E
    \fi
  \fi
\fi 

\ifnfsstwo
  \newcommand{\rmn}[1] {\mathrm{#1}}

  \DeclareMathAlphabet{\mathbfit}{OT1}{cmr}{bx}{it}
  \SetMathAlphabet\mathbfit{bold}{OT1}{cmr}{bx}{it}
  \DeclareMathAlphabet{\mathbfss}{OT1}{cmss}{bx}{n}
  \SetMathAlphabet\mathbfss{bold}{OT1}{cmss}{bx}{n}
  \ifAMStwofonts
    \ifCUPmtlplainloaded \else
      \DeclareSymbolFont{UPM}{U}{eur}{m}{n}
      \SetSymbolFont{UPM}{bold}{U}{eur}{b}{n}
      \DeclareSymbolFont{AMSa}{U}{msa}{m}{n}
      \DeclareMathSymbol{\upi}{0}{UPM}{"19}
      \DeclareMathSymbol{\umu}{0}{UPM}{"16}
      \DeclareMathSymbol{\upartial}{0}{UPM}{"40}
      \DeclareMathSymbol{\leqslant}{3}{AMSa}{"36}
      \DeclareMathSymbol{\geqslant}{3}{AMSa}{"3E}
    \fi
  \fi
\fi 

\ifCUPmtlplainloaded \else
  \ifAMStwofonts \else 
    \def\upi{\pi}
    \def\umu{\mu}
    \def\upartial{\partial}
  \fi
\fi

\begin{document}

\title{The Warp of the Galaxy and the Orientation of the LMC Orbit}

\author[Garc\'\i a-Ruiz et al.]
{I. Garc\'{\i}a-Ruiz$^1$, K. Kuijken$^1$, J. Dubinski$^2$\\
$^1$Kapteyn Astronomical Institute, Postbus 800, 9700 AV Groningen,
The Netherlands\\
$^2$Department of Astronomy and CITA, University of Toronto, 60 St. George
Street, Toronto, Ontario M5S 3H8, Canada}
\maketitle

\begin{abstract}
After studying the orientation of a warp generated by a companion
satellite, we show that the Galactic Warp would be oriented in a
different way if the Magellanic Clouds were its cause. We have
treated the problem analytically, and complemented it with numerical
N-Body simulations.  
\end{abstract}

\begin{keywords}
galaxies:kinematics and dynamics -- Galaxy:structure -- Magellanic Clouds
\end{keywords}

\section{Introduction}

The disk of the Milky Way is remarkably flat out to 10 kpc, 
where it starts to bends in
opposite directions in the southern and northern parts. The cause of it
is still a puzzle: for a review, see Binney \shortcite{b92}.

One possibility is that the Magellanic Clouds
distort the Galactic disk in the observed way. The fact
that the direction of the maximum warping lies very close to the
galactocentric longitude of the Clouds makes this hypothesis tempting.  

The problem with this scenario is that the Clouds are not massive
enough to generate the warp amplitudes that
we observe at their present distance. This was noticed by the first
researches in this field \cite{bu57,ke57}, and later by Hunter \&
Toomre (1969). A remedy which might allow this
scenario to work was to suppose that the Clouds are currently not at the
pericenter of their orbit, so that they have been much closer to the
Galactic disk in a previous epoch ($\simeq 20$ kpc is what was needed).
However, 
later work by Murai \& Fujimoto (1986) determined the orbit of the Clouds,
and proved that the Clouds are actually at their pericenter, with an
apocenter close to 100 kpc, so the problem of the small amplitude
still remains if the Clouds are to be blamed as the cause of the
Galactic warp. 

Recently a mechanism for amplifying the
effect of a satellite has been proposed by Weinberg \shortcite{w98}.
He describes a calculation in which a disk
galaxy surrounded by a dark halo is perturbed by a massive satellite,
similar to the LMC. By means of a linear perturbation analysis, he
follows the perturbation (wake) created by the satellite in the halo,
including its self-gravity. He finds that the torque exerted on the
disk is several times larger than that due directly to the satellite:
the latter is amplified because (i) the satellite-induced wake in the
halo itself exerts a torque, roughly in phase with that from the
satellite; and (ii) the wake itself further perturbs the halo,
resulting in a torque that is larger again. Under circumstances in
which the satellite orbital frequency is close to the natural
precession frequency of the disk (i.e., the warp mode frequency of
Sparke \& Casertano \shortcite{sc}), a significant amplitude can
result.  A calculation by Lynden-Bell \shortcite{lb85} of a similar
scenario gives comparable results, as does a simple model described by
Kuijken \shortcite{k97}.

In this paper, we focus on the orientation of a warp generated by a
massive orbiting satellite with less emphasis on the amplitude of the
warp. In \S2 we discuss a simple analytic model in
which the disk and halo are rigid: this establishes the baseline
response of a disk to satellite tides, and its orientation with
respect to the satellite orbital plane. As we show, this orientation
is different from that of the Galactic warp to the LMC orbit. \S3
contains a description of the N-body code used, and \S4 to \S6 the results of
our N-body simulations, showing that the orientation problem remains. In
\S7 we give our conclusions.

\section{Analytic results with a simplified model}

A simple model serves as a reference for the warp response of the disk
to an orbiting satellite.  Consider a rigid disk, embedded in a rigid
halo potential, and subjected to the potential of an orbiting
satellite. The evolution of the disk is governed by the combined
torque from halo and satellite.  A stellar or gaseous disk is floppy,
and so will warp when tilted, since it is not able to generate the
stresses that would be required to keep it flat; however the overall
re-alignment of the disk angular momentum should be comparable between
the rigid and floppy cases.

\begin{figure}
\plotone{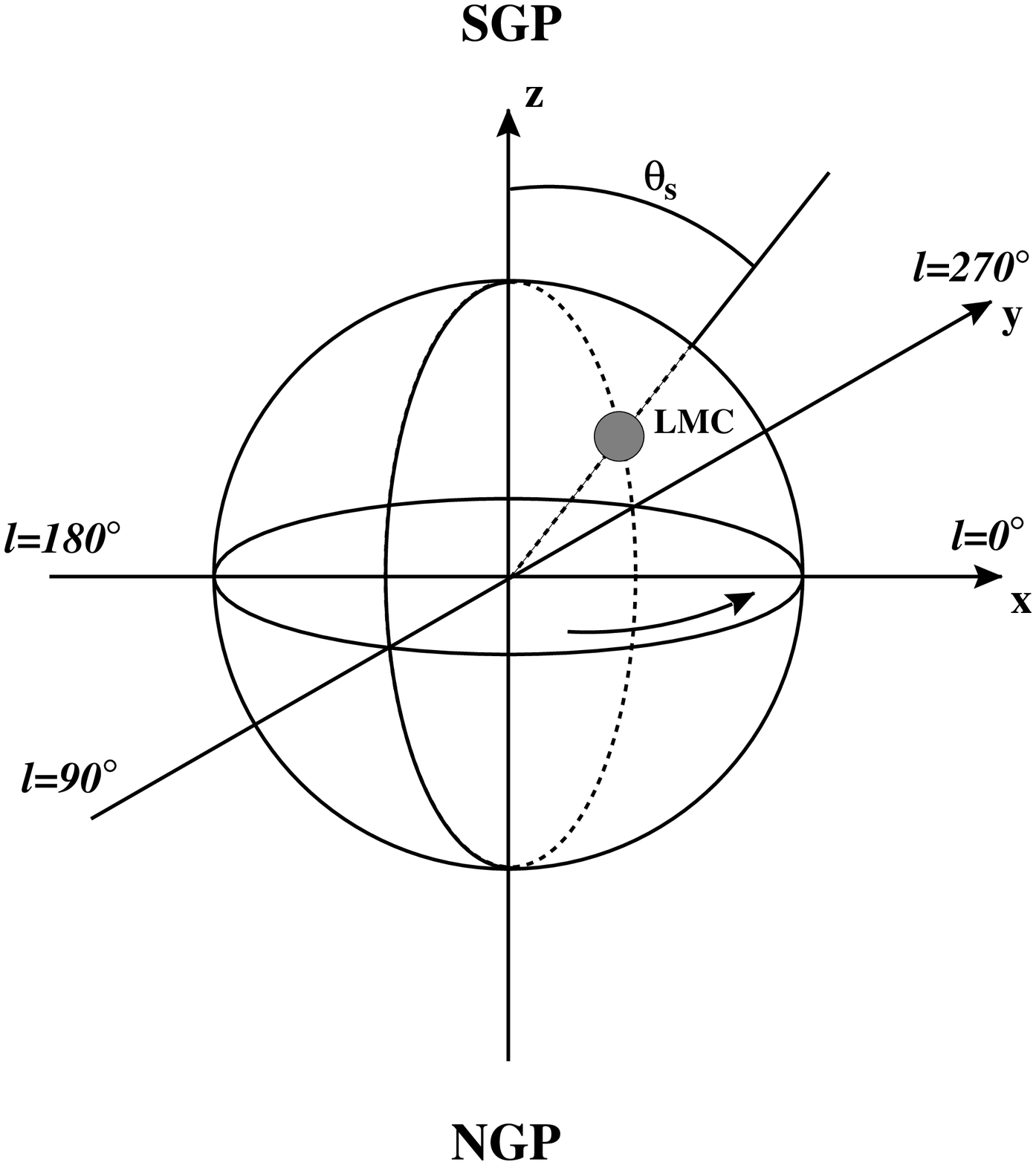}
\caption{Definition of the coordinate system, satellite angles, and
  orientation of the disk of the Galaxy. The disk lies on the $z=0$
  plane. The sun is on the left, and the galactic plane is viewed from
  the South Galactic Pole so that the disk rotates counterclockwise
  (indicated by an arrow).} 
\label{fig:coord}
\end{figure}

The angles used in this paper related to the satellite, and the
definition of our coordinate system are illustrated in
Figure~\ref{fig:coord}. The tilting of the disk is measured by the
angle between the $z$ axis and the angular momentum of the disk. The
longitude of this vector is the same as the longitude of the maximum
of the warp when looking at the disk from the North Galactic Pole.

The Lagrangian for a rigidly spinning, axisymmetric object is
\newcommand{\half}{\textstyle{\frac12}}
\begin{equation}
{\cal L}=\half I_1(\dot\theta^2 + \dot\phi^2 \sin^2\theta)
+ \half I_3(\dot\phi\cos\theta + \dot\psi)^2 
-V(\theta,\phi)
\end{equation}
where $(\theta,\phi,\psi)$ are the Euler angles, and $I_3$ and $I_1$
are the moments of inertia of the object about its symmetry axis
and about orthogonal directions. $V$ is the potential energy of the
body in the halo plus satellite potential. The $\psi$-equation of
motion leads to the conserved quantity
$S=I_3(\dot\phi\cos\theta+\dot\psi)$, the spin, and the other two
equations of motion then become
\begin{equation}
I_1\ddot\theta-I_1\dot\phi^2\sin\theta\cos\theta
+S\dot\phi\sin\theta+{\partial V\over\partial\theta}=0
\label{eq:theq}
\end{equation}
and
\begin{equation}
I_1{{\rmn d}\over{\rmn d}t}({\dot\phi\sin^2\theta})
+{\partial V\over\partial\phi}=0.
\label{eq:pheq}
\end{equation}
For small deviations from the equator ($\theta=0$), we can expand
these equations in terms of $x=\sin\theta
\cos\phi\simeq\theta\cos\phi$, $y=\sin\theta
\sin\phi\simeq\theta\sin\phi$. In these terms the equations of motion
become
\begin{equation}
I_1\ddot{x}+S \dot{y} +{\partial V\over\partial x}=0,
\end{equation}
\begin{equation}
I_1\ddot{y}-S \dot{x} +{\partial V\over\partial y}=0.
\end{equation}
For small $x,y$, the potential energy of the disk due to the flattened
halo will have the form $\half V_{\rmn H}(x^2+y^2)$, and that due to
the satellite at position $\theta_{\rmn S},\phi_{\rmn S}$ will be
$-V_{\rmn S}(\sin^2\theta_{\rmn S} - x\sin2\theta_{\rmn
  S}\cos\phi_{\rmn S} - y\sin2\theta_{\rmn S}\sin\phi_{\rmn S})$ where
$V_{\rmn H}$ and $V_{\rmn S}$ are constants representing the strengths
of the halo 
torque and of the quadrupole of the tidal field from the satellite,
respectively. Hence we find
\begin{equation} 
I_1\ddot{x}+S \dot{y} + V_{\rmn H}x  +
V_{\rmn S}\sin2\theta_{\rmn S}\cos\phi_{\rmn S}=0, 
\end{equation}
\begin{equation}
I_1\ddot{y}-S \dot{x} + V_{\rmn H}y +
V_{\rmn S}\sin2\theta_{\rmn S}\sin\phi_{\rmn S}=0.
\end{equation}
If furthermore the satellite orbit is circular and polar in the $x=0$ plane, 
$\theta_{\rmn S}=\Omega_{\rmn S}t$, $\phi_{\rmn S}=90$, 
and the solution to the equations of motion is 
\begin{equation}
x={2\Omega_{\rmn S}S\over\Delta}V_{\rmn S}\cos2\Omega_{\rmn S} t;
\quad
y={4I_1\Omega_{\rmn S}^2 - V_{\rmn H}\over\Delta}V_{\rmn S}\sin2\Omega_{\rmn S}t
\label{eq:rigidsol}
\end{equation}
plus free precession and nutation terms, where $\Delta=(V_{\rmn
H}-4I_1\Omega_{\rmn S}^2)^2-4\Omega_{\rmn S}^2S^2$. (A more general
quasiperiodic satellite orbit yields a solution which can be written
as a sum of such terms.) Notice that the satellite provokes an
elliptical precession about the halo symmetry axis, with axis ratio
dependent on the halo flattening and on the satellite orbit
frequency. For example, for an exponential disk of mass $M$, scale
length $h$ and with a flat rotation curve of amplitude $v$,
$I_3=2I_1=6Mh^2$ and $S=2hvM$. For such a disk in a spherical (or
absent) halo ($V_{\rmn H}=0$), a satellite orbiting at radius $r_{\rmn
S}$ has frequency $\Omega_{\rmn S}=v/r$, and hence the axis ratio of
the forced precession is $(x:y)=r_{\rmn S}/3h$. Hence the response of
the disk to a distant satellite is mainly to nod perpendicular to the
satellite orbit plane. This result can be understood as the classic
orthogonal response of a gyroscope to an external torque: a distant
satellite has a sufficiently low orbital frequency that the disk
responds as if the torque were static.

For a slightly flattened potential of the form $\half
v^2\ln[R^2+(z/(1-\epsilon))^2]$, $V_{\rmn H}=Mv^2\epsilon$.  With
non-zero $\epsilon$, the axis ratio of the precession cone 
becomes $[(4h/r_{\rmn S}) / (\epsilon-12h^2/r_{\rmn S}^2)]$: again the
oscillation in $x$ is larger than that in $y$ except for very
flattened halos.

\begin{figure}
\plotone{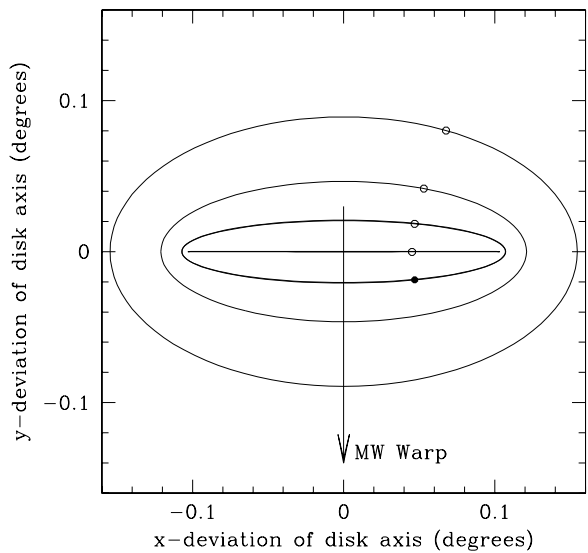}
\caption{The oscillation of the axis of a rigid exponential disk
subjected to the tidal field of an orbiting satellite. The amplitude
is calculated assuming a satellite of mass $1.5\times10^{10}M_\odot$,
orbiting at radius $54\rmn kpc$ in the $z=0$ plane. The direction of
the tilt of the Galactic disk with respect to the Magellanic Clouds'
orbital plane is indicated by the arrow. The dots mark the expected
position of the disk axis given the current phase of the LMC orbit for
(bottom to top) halo potential ellipticities
$\epsilon=0$ (solid symbol),$0.05,0.1,0.15$ and 0.2 (open circles).}
\label{fig:rigidtilt}
\end{figure}

The amplitudes generated by tidal perturbation from a satellite such
as the LMC are small, less than a degree. The largest amplitude of
oscillation is in the $y$-direction. The potential energy of the disk
due to the tidal field of the satellite can be shown to be (see
Appendix)
\begin{equation}
V_{\rmn S}={3GM_{\rmn S}I_1\over2r_{\rmn S}^3}.
\end{equation}
Hence equation~\ref{eq:rigidsol} yields, to leading order in
$h/r_{\rmn S}$, an $x$-amplitude of 
\begin{equation}
\frac98 \frac{GM_{\rmn S}}{v^2 r_{\rmn S}}\frac{h}{r_{\rmn
S}}
\simeq 0.09^\circ
\end{equation}
for the LMC (orbital radius about $50\,\rmn kpc$, and $r_{\rmn
S}/h\simeq15$). This number increases only slightly (a factor 2) for
halo flattenings up to 0.2 (see Figure~\ref{fig:rigidtilt}).

It is clear from this calculation that simple tidal tilting of a disk
by an LMC-like satellite does not provide a good model for the warping
in the Galaxy, because the orientation of the warp is not perpendicular
to the  orbital plane of the LMC.  This constraint is independent of
the strength of the perturbation $V_{\rmn S}$.

The amplitudes are much too small, but we have only considered the
tilting of a rigid disk, and the situation can change when the
floppiness of the disk is considered.

\section{Simulation details}

To test this scenario, and in particular to get beyond the rigid
tilting considered above, we have performed some N-body
simulations. We assume the halo to be a background potential which
does not respond to the disk or the satellite.

The N-body code used for this work models the disk as a set of
concentric spinning rings embedded in a spherical, rigid halo. This
description allows warps to be described, but not in-plane distortions
of the disk such as bars or lopsidedness.

\subsection{Initial conditions}

We have performed simulations with two types of disks: a rigid
disk and a exponential disk. The rigid disk run tells us how good the
analytic predictions are, and the exponential disk
is used later for a more realistic approach. 

\begin{figure}
\plotone{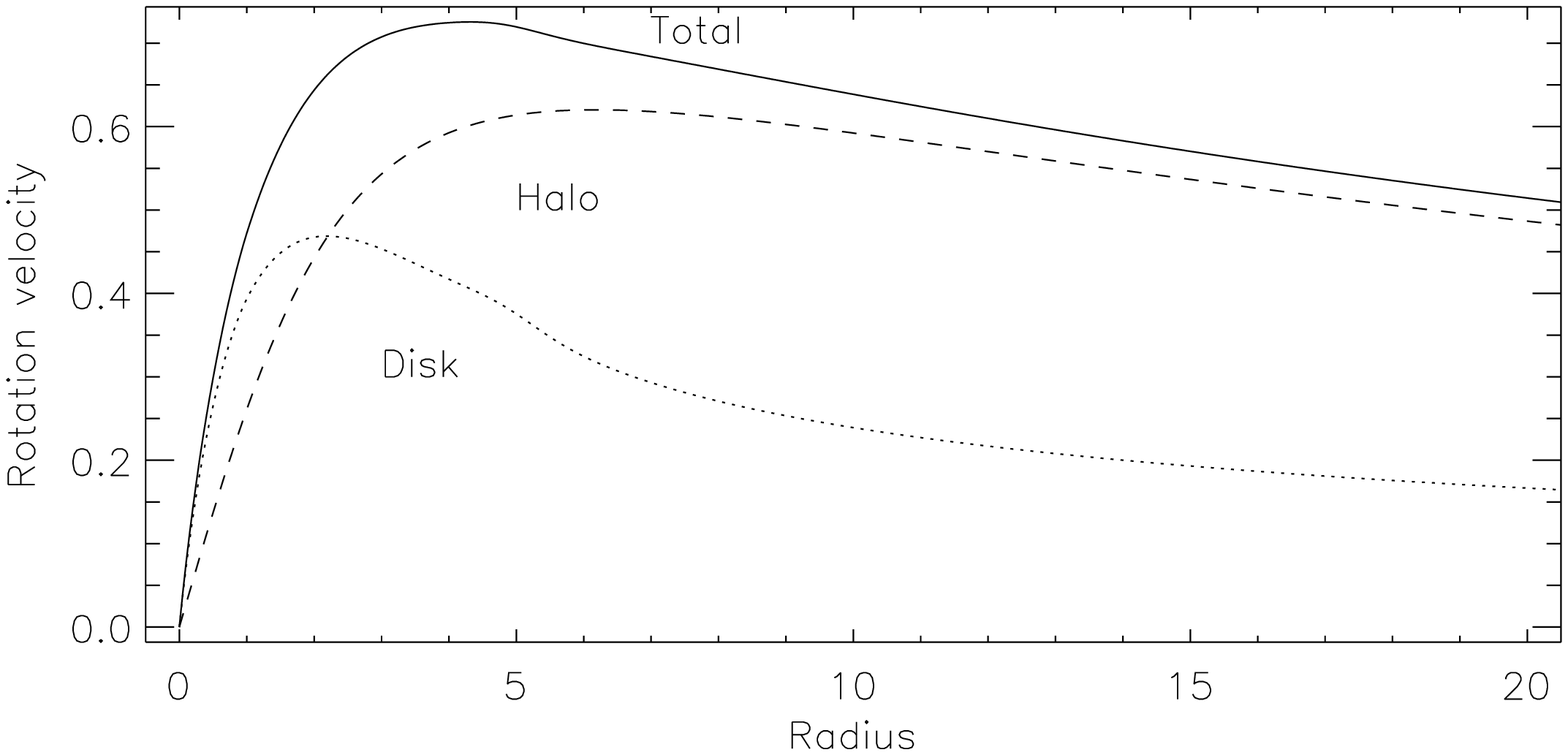}
\caption{{Rotation curve showing the contribution from the disk
    (dotted) and halo (dashed) to the total (solid)}}
\label{fig:rc}
\end{figure}

We use a King model for the halo, in order to obtain a reasonable flat
rotation curve (Figure~\ref{fig:rc}). This is accomplished with
a model of $\Psi_0/{{\sigma}_0}^2 = -6$, a tidal radius of 44 (200
kpc. for a 4.5 kpc scale-length disk), and mass of 10 disk masses.  

Each of the disks is made of 1000 rings, each of them consisting of 36
particles. Various runs where made with more rings and more particles
per ring, without important changes in the results described below.

The satellite is modelled as having a Plummer distribution. To avoid
relative movements of the galaxy with respect to the satellite we
have used two satellites instead of one, symmetrically placed with
respect to the centre of the halo-disk system. This causes the dipole
term of the tidal field to be zero, avoiding relative movements of the
galaxy with respect to the satellites. It is equivalent to only
keeping the even-$l$ 
terms in the potential of the satellite, neglecting the
dipole, $l=1$, terms in the potential, and concentrating on the warp
(which result from the quadrupole, $l=2$ terms).

The first run was made with a satellite in a circular orbit, to try to
reproduce the predictions in \S2. Later a non-circular orbit is
considered, and the difference between both simulations analysed. The
non-circular orbit has a pericenter at 50 kpc and apocenter at 100
kpc, consistent with recent determination of the orbit of the
Clouds \cite{lin}. In the non-circular simulations the satellite
starts at its apocenter at the beginning of the simulation, where the
perturbation on the disk is the smallest possible.

The units of the model translate to the Galaxy (disk scale-length
$4.5\,\rmn kpc$, and the rotation velocity at $8.5\,\rmn kpc$ of
$220\,\rmn km\,s^{-1}$) as follows: length unit = $4.5\,\rmn kpc$,
velocity unit = 340 $\rmn km\,s^{-1}$, time unit = $1.30 \times
10^7\,$years, mass unit= $1.20 \times 10^{11}\rmn M_{\odot}$.  With
these numbers, the disk mass of our model is $6.1 \times 10^{10}\rmn\,
M_{\odot}$, and the satellite (LMC) has a mass of $1.5 \times 10^{10}\rmn\,
M_{\odot}$, the biggest current mass estimate for the Clouds
\cite{so92}. 
In the coordinate system of the simulations, the $z=0$ is the disk
plane, and the orbit of the satellite lies in the $x=0$ plane.

\subsection{Code used to evolve the system}

The disk is modelled as a system of pivoted spinning rings, fixed at
the centre of the halo. Each ring is realized as 36 azimuthally-spaced
particles, and the potential generated by the rings is calculated with
a tree code \cite{bh86}. The forces on the individual
``ring-particles'' are used to calculate the torque on each ring.  The
force exerted by a satellite on the ring particles was evaluated
directly using the Plummer law. 

The Euler equations for rings and axisymmetric bodies can be rewritten in a
form so that the time derivatives of the instantaneous angular velocities
about the body axes are linear combinations of the angular velocities, 
torques and body normal vector components.  This allows the derivation of a
second order explicit time-centred leapfrog integration scheme that
can be used to solve the coupled equations for the rings and make it easy
to merge with an N-body code \cite{du00}.

\section{Rigid Disk}

\begin{figure}
\plotone{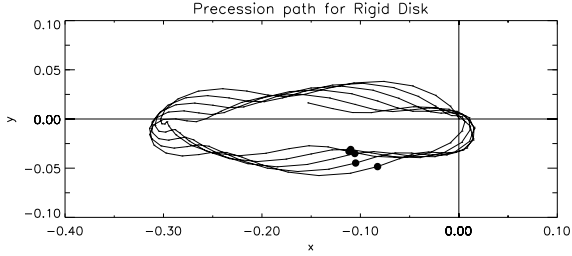}
\caption{Precession path followed by the rigid disk:
  $x=\theta\cos\phi$, $y=\theta\sin\phi$. The dots indicate the disk's
  state when the satellite has the LMC's orbital phase. } 
\label{fig:rigidrun}
\end{figure}

As a first approach, we have evolved a rigid disk and analysed its
evolution under the influence of an orbiting satellite. The result of
our simulation is in good agreement with the analytic
predictions. The disk wobbles under the influence of the satellite,
describing an ellipse elongated in the direction perpendicular
to the satellite's orbital plane. The path followed by the disk is
plotted in Figure~\ref{fig:rigidrun}, where it can be seen that most
of the time the maximum of the warp is located in the direction
perpendicular to the orbital plane of the satellite ($l=0^\circ$ and
$l=180^\circ$). The ellipse isn't as regular as in
Figure~\ref{fig:rigidtilt} for two reasons: the assumption that the disk is
much smaller than the orbital radius of the satellite is not
completely fulfilled; and there are some transient terms present
because of the initial conditions of the simulation. This are also the
cause for the precession ellipse of the disk not to be centred in the
origin. 

The position of the warp when the satellite is at the location of the
LMC is indicated by the dots in Figure~\ref{fig:rigidrun}, and the
location of them resembles the predicted one in
Figure~\ref{fig:rigidtilt} (for $\epsilon_{halo}=0$) remarkably well.

\section{Exponential self-gravitating Disk}

We now consider a more realistic disk: an exponential disk model, in
which we have considered also the disk's self-gravity. The first thing
that draws our attention in this simulation is a peak we see in the
inclination at around 6.5 scale-lengths. Simulations done with a
different rotation curve showed that this peak occurs at the locations
on the disk that satisfy $\Omega_s/w_z=2,3,...$, that are caused by
resonances with the satellite's orbital frequency. This happens
because the non-linear behaviour of the outer parts of the disk, where
the assumption $r_s \gg r_{disk}$ is worse than it is further in.

This is not the kind of warp we are looking for, due to the fact that
it is the result of a satellite with a single frequency, and in the
real case the eccentric orbit of the satellite will wash out this
peak. Looking at the evolution of the disk it is clear that the warp
looses its coherence at a radius about 4.5 scale-lengths (at larger
radius the line of nodes winds up), so we will measure the warp
properties considering that the disk finishes there. 
 
\begin{figure*}
\plotnine{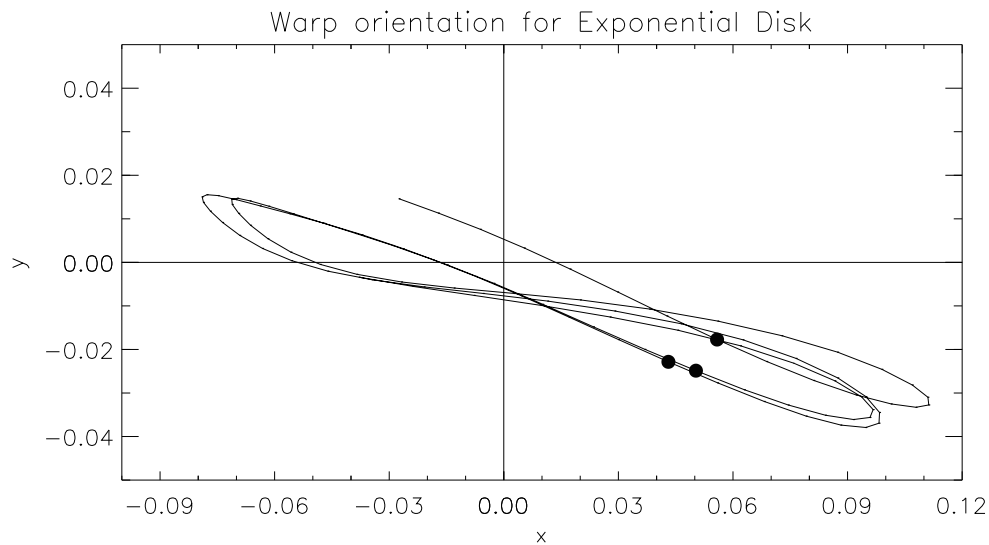}
\caption{Warp orientation
  followed by the Exponential Disk:
  $x=\theta\cos\phi$, $y=\theta\sin\phi$. The dots indicate the disk's
  state when the satellite has the LMC's orbital phase.}
\label{fig:exprun}
\end{figure*}

In the case of a floppy disk it is not 
straight-forward to define a single inclination and position
angle. We have separated the disk into two components: the inner disk
and the outer (warped) disk. The inner disk consists on the first 2
scale-lengths, and remains practically flat along the simulation. The
warping angle is then calculated as the angle between the inner and
outer disk vectors. We have chosen to use the disk vectors and not
the angular momentum, for example, not to penalise the outer less
massive rings. The results presented here do not change significantly
when the definition of the inner disk is altered.

It has to be kept in mind that the warping angles quoted here are
different than the maximum amplitude of the warp, who usually are
larger by a factor not greater than 5.

Using this method we obtain a plot similar to
Figure~\ref{fig:rigidrun} for the exponential disk, which is
shown in Figure~\ref{fig:exprun}. Only the path after t=160 is shown,
that is the moment when the disk behaviour reaches an equilibrium.

Note that the predictions for the Galactic Warp don't really change
with the floppiness of the disk: it is clearly close to $l=0^\circ$,
as chapter \S2 predicted, and not at $l=\simeq 90^\circ$, as we
observe it in the Galaxy. 

\section{Non-circular orbit, and flattened halos}

We also considered non-circular orbits, to allow for the fact that the
orbit derived for the Clouds has a pericenter of 50 kpc and an
apocenter of 100 kpc \cite{mf80}. The changing radius of the satellite causes a
fluctuating tidal field amplitude, which could be important for the
dynamics of the disk. Here we show that in fact the effect does not
change our conclusions materially.

First, to have an idea of what to expect, we integrated the analytic
equations of section \S2 with a satellite in this kind of orbit. The
result was, as before, that the disk's precession path was contained
within an ellipse, elongated along the direction perpendicular to the
satellite's plane. This causes the warp maxima to be most of the time
close to the direction perpendicular to the satellite's orbit.

We then performed simulations with this type of orbits. The first thing we
observe in these simulations is that the resonance peak we
found in the circular orbit simulation has disappeared. Now the
satellite doesn't have a single frequency, so the result is not
surprising. The energy of the resonance gets distributed along
different parts of the disk now, and no coherent pattern can be
maintained across the disk, winding up the outer parts of the
disk. When we look at the inner 4.5 scale-lengths as before, the
precession pattern remains similar to the simulation with the circular
orbit, so does the prediction of the warp's longitude at LMC's actual
orbital phase. So our conclusions are not modified by the
non-circularity of the orbit.

The halos considered in all these simulations are spherical, which
means that they don't contribute to the generation of torques on the
disk. We know that halos are not spherical, which creates a
preferred plane in which the disk settles. Ellipticities of the order
of 0.05 in the potential make the precession paths described before yet
more elongated, which would make the chances of finding the warp
maxima in the satellites' direction even more unlikely.

\section{Conclusion}

We show by means of analytic work and N-body simulations that the
precession path of a warp generate by an orbiting satellite galaxy is
elongated along the direction perpendicular to the satellite's orbital
plane. Applying our result to the Milky Way, if the Galactic Warp is
generated by the Magellanic Clouds, the direction of maximum amplitude
of the warp would lie close to $l \simeq 0^\circ$, as compared to
the observed direction of $\simeq 90^\circ$. Even if the halo's
self-gravitating tidal response to the satellite amplifies the effect
of the satellite~\cite{w98}, this response will be mostly in phase
with the satellite, and the alignment problems will persist. Possibly
the Sagittarius dwarf galaxy, whose orbit lies at 90$^\circ$ to that
of the LMC, can be the cause of the warp instead~\cite{ibata}.   

A limitation of the present work is that the halo has not been
considered as a live, self-gravitating component. It has been
shown~\cite{dk,nt} that the back-reaction of the halo on a
re-aligning disk can have important consequences. Such effects will be
the subject of a further paper.

\appendix

\section{Potential of axisymmetric disk due to a satellite}

The potential energy of an disk of surface density $\Sigma(r)$ and in
the gravitational field due to a satellite at position ${\bmath
r}_{\rmn S}$ is given by
\begin{equation}
V=-\int {\rmn d}^2{\bmath r}\,G\Sigma(r)
{M_{\rmn S}\over|{\bmath r}-{\bmath r}_{\rmn S}|}.
\end{equation}
Choosing spherical coordinates for the satellite's position (see
Figure~\ref{fig:coord}), and Cartesian coordinates in the disk plane so
that the satellite has $x=0$, we have 
\begin{equation}
V=-GM_{\rmn S}\int \Sigma{\rmn d}x\,{\rmn d}y\,
(r_{\rmn S}^2-2yr_{\rmn S}\sin\theta_{\rmn S}+x^2+y^2)^{-1/2}.
\end{equation}
Assuming that the disk is small compared to $r_{\rmn S}$, we can
expand the integrand in $x$ and $y$. For an axisymmetric disk the
second-order terms are the first ones that generate a potential
gradient: they are
\begin{equation}
V=-{GM_{\rmn S}\over r_{\rmn S}^3}\int\Sigma{\rmn d}x\,{\rmn d}y\,
[-\half(1-3\sin^2\theta_{\rmn S}) y^2-\half x^2]
\end{equation}
which results in 
\begin{equation}
V=-{3GM_{\rmn S}I_1\over 2r_{\rmn S}^3}\sin^2\theta_{\rmn S}+\hbox{constant}.
\end{equation}

\newcommand{\apj}{ApJ}
\newcommand{\mnras}{MNRAS}

\end{document}